\documentclass[jkps,fleqn,showpacs,showkeys,twocolumn]{revtex4}
\usepackage{graphicx}
\usepackage{amssymb}
\usepackage{amsmath}
\usepackage{bm}

\usepackage{multirow}
\usepackage{tabularx}

\usepackage{verbatim}

\begin{document}
\setcounter{page}{0}
\title[]{Tracing the evolution of physics with a keyword co-occurrence network}
\author{Taekho \surname{You}}
\affiliation{Department of Industrial and Management Engineering, Pohang University of Science and Technology 37673}
\author{Oh-Hyun \surname{Kwon}}
\affiliation{Department of Physics, Pohang University of Science and Technology 37673}
\author{Jisung \surname{Yoon}}
\affiliation{Department of Industrial and Management Engineering, Pohang University of Science and Technology 37673}
\author{Woo-Sung \surname{Jung}}
\email{wsjung@postech.ac.kr}
\affiliation{Department of Industrial and Management Engineering, Pohang University of Science and Technology 37673}
\affiliation{Department of Physics, Pohang University of Science and Technology 37673}

\date[]{}

\begin{abstract}
Describing the evolution of science is a salient work not only for revealing the scientific trend but also for establishing a scientific classification system. In this paper, we investigate the evolution of science by observing the structure and change of keyword co-occurrence networks. Starting from seven target physics fields and their initial keywords selected by experts from the Korean Physical Society, we generate keyword co-occurrence networks better to capture topological structure with our proposed approach. In this way, we can construct a more relevant and abundant keyword network from a small set of initial keywords. With these networks, we successfully identify the scientific sub-field by detecting communities and extracting core keywords of each community. Furthermore, we trace the temporal evolution of sub-fields with the time-snapshot keyword network, the resultant temporal change of the community membership explains the evolution of the research field well. Our approach for tracing the evolution of the research field with a keyword co-occurrence network can shed light on identifying and assessing the evolution of science.

\end{abstract}



\maketitle

\section{Introduction}

With the increasing complexity of science, understanding the evolution of science and assessing scientific research have become significant issues in allocating grants to increase the efficiency of research performance\cite{Shibayama11Sciento, Aagaard20QSS}. The growth of interdisciplinary research and the accelerating changes in scientific development trends make the problem difficult as well. To overcome these problems, scholars tried to describe the shape of science by mapping publications. In bibliometric research, they try to provide a way to describe scientific knowledge with publications and propose how to increase efficiency in research funding\cite{Abramo09Respol}. Furthermore, as data accessibility increases, the complex network has been emerging as a promising tool that can provide rich intuition on the knowledge structure\cite{Foster15ASR, Kuhn14RPX}.

From the other point of view, understating the evolution of science is also crucial in establishing a scientific classification system, since the assessment of research output and distributing grants highly depends on the classification system. For the purpose, conventional classification systems have been invented, {\it e.g.,} ASJC, PACS, and PhySH show several practical results\cite{Bornmann10Pone, Radicchi11Pre}. Still, conventional classification systems are manually designed by a small group of experts, and it is too slow to catch up with the current scientific development trend. To overcome those limitations, a bottom-up way to define a classification system that captures fast-changing research fields leveraging collective intelligence has been proposed \cite{Yoon18Saemulli}.

Then, how can we identify the evolution of science quantitatively? Before answering this question, we need to understand the structure of scientific knowledge first. The keyword co-occurrence network is a popular method to capture the knowledge structure~\cite{Liu17Pone, Liu18arxiv, Rad17Pone, Su10Sciento, Kim2016Ant, Choi11JIIS, Yi12Sciento}. After the network construction, centrality metrics such as betweenness centrality and eigenvector centrality are commonly used to identify the importance of the keyword, or network visualization is also used to classify the relationship between keywords. In this way, the topological structure of the science can be extracted systemically.

However, how to construct the network affects the quality of downstream tasks. The most popular way to construct a network is by collecting all keywords in the target document and removing the keywords below the threshold. This \textit{bottom-up} approach is the simplest way, but the constructed network tends to possess many irrelevant keywords. On the other side, \textit{top-down} approach, which constructs a network with keywords that manually selected by experts, is also possible. With this approach, the network is small enough to look into and less likely to have noise, while it requires domain knowledge and a long time to select the keywords. Furthermore, it can be subjective, because a small group of experts selects keywords. Both approaches have clear pros and cons, and it is necessary to develop a method that can leverage the advantages of both.

In this paper, we attempt to capture the evolution of the physics field with scientific publications with keywords of each publication. For the purpose, we propose \textit{coarse-grained} approach to construct an improved keyword occurrence network, encompassing both \textit{bottom-up} and \textit{top-down} approaches. With the constructed network, we identify the research field with a community detection method, named the research field systematically. The extracted network and the detected research field are reasonable and well-matched with projects done by experts from the Korean Physical Society (KPS). With this network, we track the evolution of the detected research field across different time snap-shot networks and successfully capture the evolution of the physics field.

\section{Data \& Method}

\begin{figure*}
    \includegraphics[width=\textwidth]{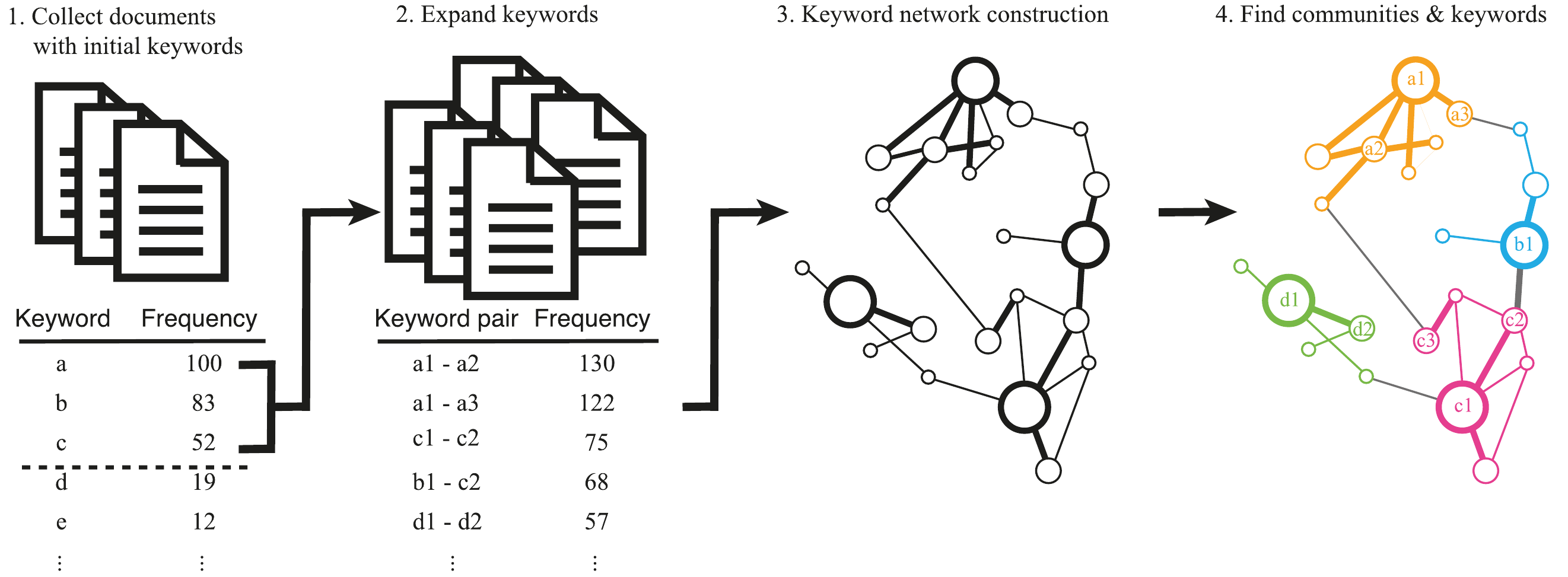}
  \caption{\textbf{Schematic diagram for generating a keyword co-occurrence network.} Starting from the initial keyword set, we source documents from WoS. From the collected documents, we calculate the frequency of each keyword, and frequent keywords are added to the keyword set. With this expanded keyword set, we re-collect documents and construct keyword network with co-occurrence of keyword pairs. We prune the network by removing links with weight below the threshold $\alpha$. Finally, we apply the community detection method and compute centrality for identifying sub-discipline.}
    \label{fig_schematic}
\end{figure*}

We source 1,359,750 publications from Web of Science, which are classified as physics disciplines, and published from Jan. 2011 to Dec. 2018. Each publication contains keywords that are provided by authors. These user-generated keywords have various variations, so we manually standardize the plural form, present continuous tense, an abbreviation of the singular form, and remove symbols such as a hyphen. 

Changing the plural form to singular form may affect the analysis, but the results were robust. For example, the keywords \textit{quantum dot} and \textit{quantum dots} are located in a different position in the network. \textit{Quantum dots} tend to connect with theoretical words while \textit{quantum dots} connect with materials. Removing the plural form would cause the merger of two different communities, but we check that these two keywords are always classified as the same community. It implies that changing plural forms into singular forms does not affect the following analysis even in the exceptional case.

We construct a keyword network with a \textit{coarse-grained} manner. First, we start from a small set of initial keywords which are selected by domain experts. This initial keyword set is usually fundamental keywords that can represent the target research field (Table~\ref{tab:statistics}). Then, we collect the documents, which contain these keywords, and rank all keywords by frequency in the collected documents except for the initial keyword set. Next, add the keywords to the keywords set up to the top 30. Second, we recollect the documents with updated keywords set and construct a weighted keyword co-occurrence network whose node is a keyword and the weight of the link between two nodes is the number of publications containing both keywords. In this way, we can obtain the keyword network that contains not only initial keywords but also relevant keywords in a systematic way.

The constructed network could have irrelevant keywords. To eliminate these irrelevant keywords, we apply the threshold of $\alpha$. A link below the threshold is removed, and a node is also removed if it has no links above the threshold. This pruned network can have multiple connected components, and we used the largest connected component for the result. Finally, we apply Louvain method\cite{Blondel08JSMTE} with python package \texttt{python-louvain} to find the research field in the keyword network. Louvain method has a resolution parameters to control the size of the result communities. For simplicity, we  and we set the parameter value as a default value of 1. We define each community as an extracted research field and name each community with core keyword, which is the keyword with the highest centrality value in each community. In this work, we use degree centrality.

\section{Results}

\subsection{Research fields in Keyword Network}
\label{a}

We apply our method to seven physics fields, which are selected based on the KPS report to describe the evolution of emerging academic fields. Target physics fields and their initial keywords are listed in Table~\ref{tab:statistics}. We have two fields that have single initial keywords. Other fields have multiple keywords since they have multiple keywords expressing their fields. Using these keywords, we collect publications and build a keyword co-occurrence network.

The basic statistics of the publications and generated networks are also summarized in Table~\ref{tab:statistics}. We collect more than 20,000 publications except for \textit{the origin of the universe} field. The number of publications in \textit{the origin of the universe} field is small because our dataset covers only publications in the physics field, not in the astronomy field of WoS. Because of this, the network would explain the structure in the physics field. We also see that about 99\% of keywords are excluded when we construct the network. It means most of the keywords appear less than 5 times with other keywords. For the remaining keywords, we find communities and core keywords with high centrality.

\begin{table*}
    \centering
    \caption{\textbf{Initial keywords and basic statistics for keyword co-occurrence network generation.} After generating a network, nodes which have no links over $\alpha = 5$ are removed.}
    \begin{ruledtabular}
        \begin{tabular}{c|c|cccc}
        Physics field & Initial keywords  & \begin{tabular}{c}Number of\\publications\end{tabular} & \begin{tabular}{c}Number of\\unique keywords\end{tabular} & \begin{tabular}{c} Keywords in \\ network\end{tabular} & \begin{tabular}{c}Number of\\ communities\end{tabular}\\
        \colrule
        Complex system & complex system & 22,958 & 51,924 & 581 & 7 \\
        \colrule
        Biophysics & biophysics & 47,712 & 86,004 & 714 & 8 \\
        \colrule
        \begin{tabular}{c} Quantum \\ computing \end{tabular} & \begin{tabular}{c}quantum computing, quantum computer\\quantum computation, qubit \end{tabular}& 33,744 & 64,976 & 546 & 7\\
        \colrule
        Quantum matters & \begin{tabular}{c}quantum matter, quantum many body,\\ quantum material,
        quantum magnet,\\ electron spectroscopy, complex material,\\
        correlated system, topological insulator,\\ many body effect\end{tabular} & 68,979 & 90,135 & 592 & 9\\
        \colrule
        \begin{tabular}{c}The origin\\ of the universe \end{tabular}& \begin{tabular}{c}standard model, string theory, rare isotope\\
        dark matter, astrophysics\end{tabular} & 7,583 & 14,112 & 729 & 9\\
        \colrule
        AI \& big data & \begin{tabular}{c}neural network, artificial intelligence,\\ big data, machine learning\end{tabular} & 93,692 & 170,931 & 603 & 8 \\
        \colrule
        Energy & \begin{tabular}{c}fusion energy, renewable energy, \\ neutron matter, neutron source,\\ future energy \end{tabular} & 41,058 & 75,038 & 533 & 10\\
        \end{tabular}
        \end{ruledtabular}
    \label{tab:statistics}
\end{table*}
    
\begin{figure*}
    \includegraphics[width=\textwidth]{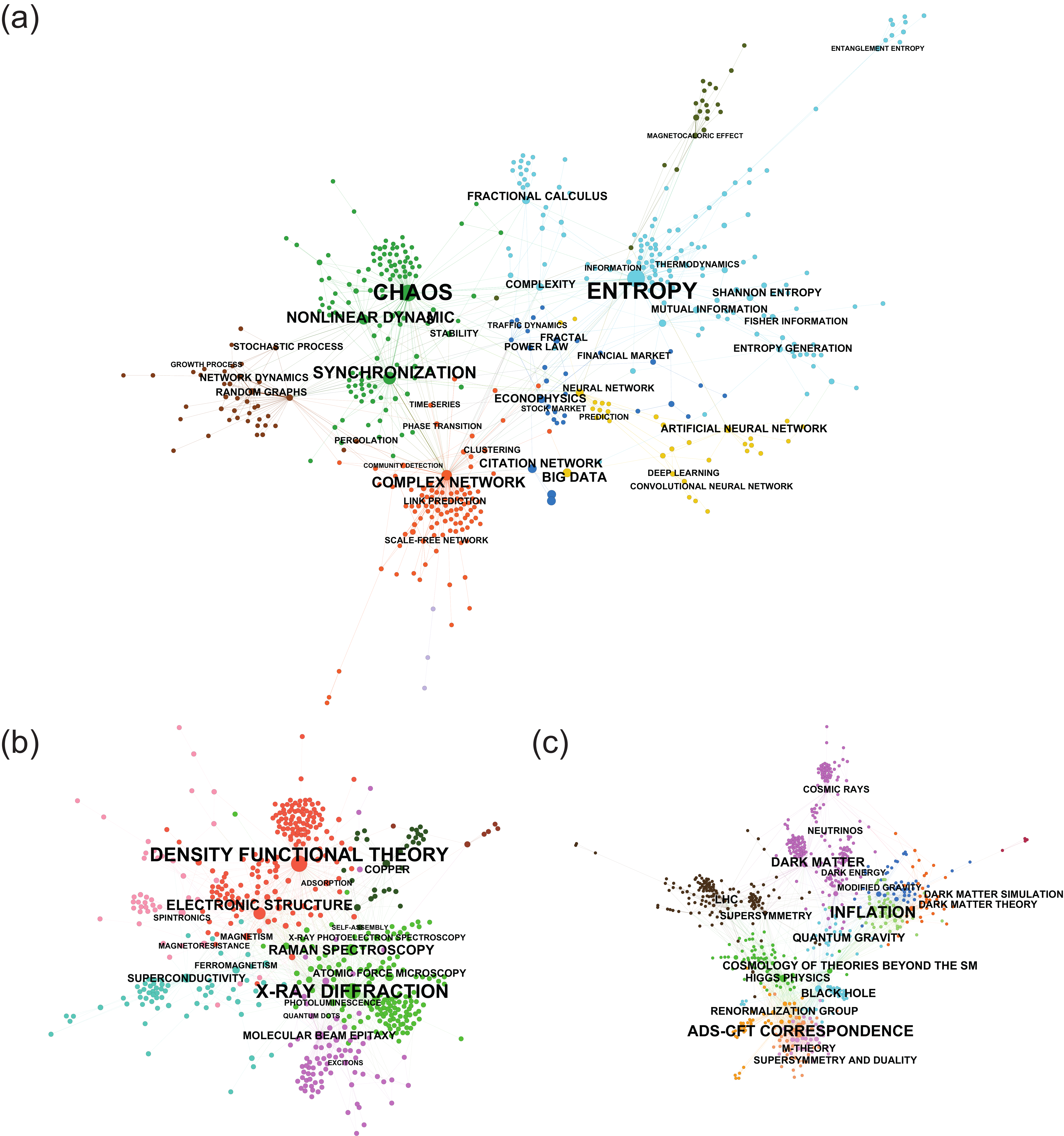}
    \caption{\textbf{Keyword co-occurrence network with field (a) complex system, (b) quantum matter, and (c) the origin of the universe.} Each Node represents a keyword, the two nodes are connected if they co-occur in the same document, and the weight of the link indicates the number of co-occurrence across the whole dataset. Color represents the community and the size of the node is proportional to the number of appearances. In each community, keywords with the highest centrality in the community are stated. We use force-atlas 2 \cite{Jacomy14POne} for the layout.}
    \label{fig_keyword}
\end{figure*}

\begin{figure*}
    \includegraphics[width=\textwidth]{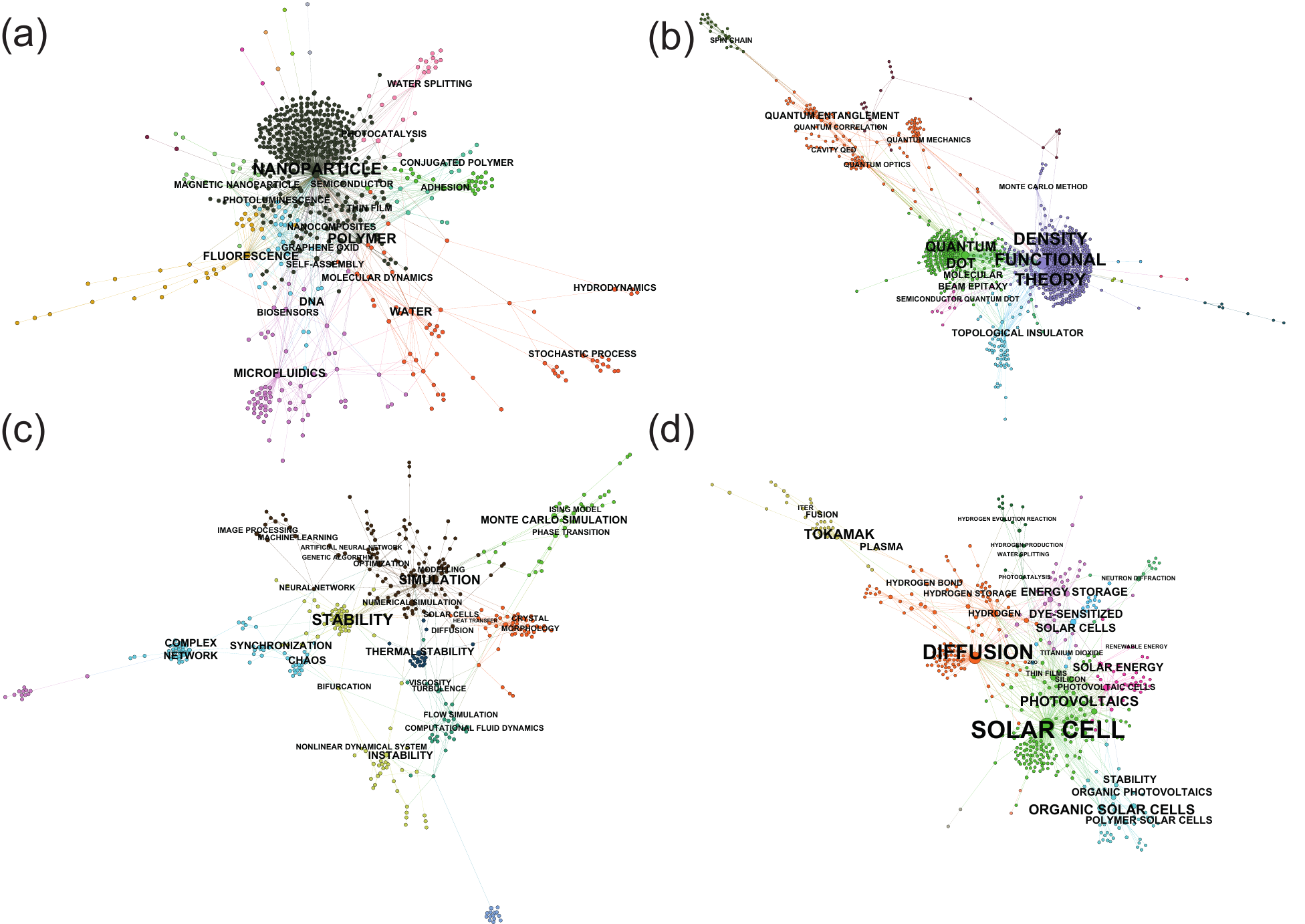}
    \caption{\textbf{Keyword network of (a) Biophysics, (b) Quantum computing, (c) AI \& big data, (d) Energy.} Same as in Figure~\ref{fig_keyword}, nodes represent keywords, and color of nodes indicates result community. }
    \label{fig_additional}
\end{figure*}

The network structures of \textit{complex system}, \textit{quantum matter}, and \textit{the origin of the universe} field are presented in Figure~\ref{fig_keyword}. And the networks of the rest four fields are described in Figure~\ref{fig_additional}. We denoted the community structure and significant nodes in the network. For example, in Figure~\ref{fig_keyword}(a), the two keywords \textit{chaos} and \textit{synchronization} consist of large communities and have high centrality values. One interesting point about core keywords is that initial keywords rarely appear as the core keywords. Due to the method to expand the keywords set, we can exhibit other relevant keywords rather than initial keywords.

The communities in the network are shown with the color in Figure~\ref{fig_keyword}. We can find 7 to 10 communities, and each community represents a subdiscipline of the target field. We can infer the characteristics of the subdiscipline by observing keywords and their connecting keywords. Some nodes are connected with a large number of other keywords, such as \textit{nanoparticle}, \textit{quantum dot}, and \textit{solar cell} (Figure~\ref{fig_additional}(a-b,d)). These connecting keywords are materials like \textit{SiO$_2$}, which implicates that these communities are dealing with various molecules and materials as their research works.

In addition to this, we can find the difference between the two communities by observing the keywords. For example, in Figure~\ref{fig_keyword}(c), we can find two communities related to dark matter. One contains \textit{neutrinos}, \textit{cosmic rays}, and \textit{dark energy}, while the other contains \textit{dark matter theory} and \textit{dark matter simulation}. Although the two communities are related to dark matter, we can infer that the first community is dealing with components while the second community is related to the theoretical approach. In summary, by observing the keywords and their connections, we can classify the characteristics of the subdiscipline and compare the difference between subdisciplines.

Then, does the keyword network map the structure of the target field? We observe \textit{complex system} and \textit{the origin of universe} network to check the details of the field. First, keywords in the field of \textit{complex systems} are clustered by 7 communities (Figure~\ref{fig_keyword}(a)). The main communities are nonlinear dynamics and information theory, which are studied theoretically in the complex system field. Keywords in these communities are \textit{chaos}, \textit{synchronization}, \textit{nonlinear dynamics}, \textit{entropy}, \textit{mutual information}, and etc. These keywords are the concepts that use in the complex system. Other communities are the applications that are studied recently in the complex system field. For example, \textit{econophysics} is connected with \textit{nonlinear dynamics}, \textit{entropy}, and \textit{complex network}. These communities have weak connections with theoretical subfields. The last community is a neural network, which is a recently emerging field. Several interdisciplinary researches have been done in this field. 

For another example, the keywords for the study of the origin of the universe are clustered by 9 communities (Figure~\ref{fig_keyword}(c)). Different from the complex system, we can see that some communities are located closely, although they belong to different communities. These communities are aggregated to two main fields: particle physics and cosmology. The community of particle physics consists of \textit{ADS-CFT correspondence}, \textit{M-theory}, \textit{black hole}, and \textit{super-symmetry}. Another one is a community of cosmology, which consists of \textit{dark matter theory}, \textit{modified gravity}, \textit{inflation}, and \textit{search for dark matter and energy}. In the middle of these two fields, there is a community that connects two large communities. \textit{Cosmology of theories beyond the standard model} and \textit{black hole} has many links to both, implying that these keywords play a crucial role to connect fields.

In summary, the keyword co-occurrence network can provide three aspects of the target field. First, the community shows which research works have been done in this field, and the keywords describe the community. Second, the target network can be described with a proper value of the parameter. The community detection method provides a resolution parameter\cite{Blondel08JSMTE}, which controls the size and number of the communities in the network. Although there exists an optimal parameter to divide the network, a proper level of the parameter can describe the description of the target field. Last, links between communities provide a clue to the evolution of the target field. As seen in Figure~\ref{fig_keyword}(c), there exists a community that connects two different fields, by observing a large number of links from the two fields. However, it does not imply a causal relationship or evolution directly.

\subsection{Tracing temporal evolution}

\begin{figure*}
    \includegraphics[width=\textwidth]{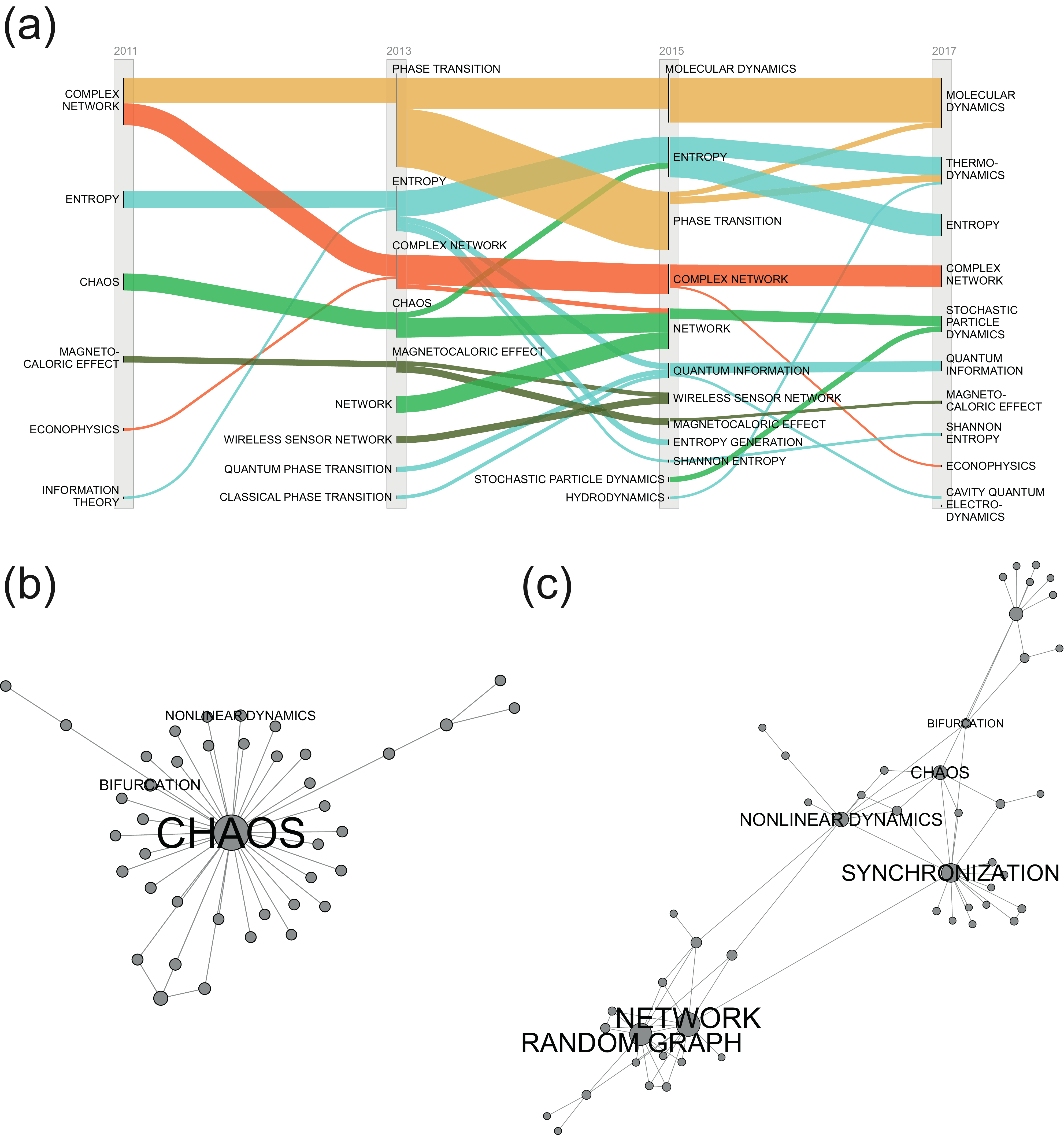}
    \caption{\textbf{Tracing temporal evolution of research field} (a) Evolution of communities in complex system field. Jaccard index is used to match the most relevant communities between two different time-snapshot networks. The thickness of the line between two communities means the number of keywords both communities share. The name of the community is the most frequent keywords in the community. (b-c) The sub-network of chaos community in 2010 and 2015 respectively. The size of the node is appearances of the keyword.}
    \label{fig_alluvial}
\end{figure*}

As the keyword network provides limited information about the evolution of the target field, we check the change of keywords in time to capture the evolution. Above the seven fields, we display the complex system field to observe this change. We construct the keyword network from the publications of three years time window and two years moving window. In detail, we construct a keyword network for a year $y$ with publications till year $y+2$ and next construct the keyword network for a year $y+2$. We adopt a three-year time window to reduce the fluctuation of keywords and a two-year moving window to avoid publication overlap between the two networks. We repeat the process to generate communities as described in section~\ref{a}

After constructing the keyword networks, we find the flow of communities within two networks. To classify the most relevant communities, we compute similarity. If a community's keywords in year $y$ mostly appear in a community in year $y+2$, we consider the two communities are most relevant. It can be described as $S_{ij} = |K_i \cap K_j|/|K_i \cup K_j|$ for the keyword set $K$ in the target community $i$ in year $y$ and community $j$ in year $y+2$. The equation is equivalent to the Jaccard similarity between two communities. In the case that the similarity $S_{ij}$ is the highest for community $i$, we consider that the community maintains its research subfield in the target field by time. Additionally, we can set a threshold $\delta$ so that the two communities are similar if $S_{ij} \ge \delta$. In this case, we can observe the merge and split of a community. In this work, we set $\delta = 0.1$.

The flow of communities in the complex system field is described in Figure~\ref{fig_alluvial}(a). The keyword mostly appears in the community, and the flow is the shared keywords between two communities. Starting from 2011, three communities of \textit{complex networks}, \textit{entropy}, and \textit{chaos} keep remaining in the largest communities for all periods. These communities are also the largest in Figure~\ref{fig_keyword}(a), supporting that these communities are the main subfields in the complex system field for all periods.

However, two significant changes appear. First, the complex network community separates into the communities of phase transition and complex network. Phase transition community becomes the largest community after the appearance and changes in the molecular dynamics community. Phase transition is an important field in statistical physics, and it becomes a large community in the complex system field as it combines with molecular dynamics. The complex network field emerges in the late 1990s, and phase transition is also studied in the complex network field. 

Second, the dominant keyword is changed in the chaos community. \textit{Chaos} is the most appeared keyword in 2011 and 2013, but the number of appearances is decreased. In 2015, a keyword \textit{network} has become the dominant keyword in the community. Note that the keyword \textit{network} is different from \textit{complex network}. It does not have any connection to the other keywords of \textit{complex network}. The change of chaos community is described in Figure~\ref{fig_alluvial}(b) and (c). It shows that the number of the appearance of keywords \textit{nonlinear dynamics} and \textit{synchronization} is increasing, while \textit{chaos} is decreasing. It means that although the research subfield keeps a remaining position in the target field, the detailed work is slightly changing over time. These changes, however, are not directly traced in Figure~\ref{fig_keyword}(a). It only shows the big community of chaos. We also see that the increase of \textit{nonlinear dynamics} and \textit{synchronization} is also related to the increment of the \textit{network}.

We also see the other minor communities in Figure~\ref{fig_alluvial}(a). These communities are merged into and split from the large community. Communities can be minor research subfields in the complex system field, or they are just a new concept and an application of existing research works. Some communities are new concepts that arose from the existing research subfields so that they are merged into the largest community. And some communities are applications of the existing subfields so that they split from the largest community.

\section{Summary}

In this paper, we inspect the evolution of physics by observing the changes of keywords. We construct a keyword co-occurrence network with a method of expanding keywords and detecting communities and their changes over time. The proposed method produces the structure of the target field, the communities of the research subfields, and their representative keywords. We apply the method to seven physics fields. The result describes the structure of the target field, and from this we can successfully figure out the scientific subfield.

Furthermore, our approach, coarse-grained manner of collecting publications, can reduce the size effect of publications. Eliminating the size effect of publications is an important work to reveal research fields. Various characteristics of the research field such as changing speed or number of researchers in the field can induce the size effect, occupying a large fraction of the network. This size effect can produce the bias of the network when we compare the two different fields. Our approach collects more relevant publications recursively starting from a small set of the initial fine-grained keywords, producing a high-quality topological map of a given field amending the size effect.

While the whole year's keyword network shows a static map of research fields, temporal evolution reveals the changes in keywords in and between the communities. Leveraging the time-snapshot keyword network, we analyze the evolution of keywords in the complex system field to understand the evolution of the subfield. We find that the largest communities remain in temporal evolution, but the core keyword in the community changes. The temporal evolution also shows the split of the community can be understood as the increase of the application field.

Despite its promising results, our approach has several limitations. First, even if two keywords refer to an identical research field  (e.g. synonym),  they can be described as separate fields in the network, resulting in different communities. This limitation can be addressed by preprocessing keywords using natural language processing techniques. Second, the proper community detection method should be considered depending on the characteristics of the network. In this work, we use the Louvain method which prefers homogeneous size and large communities. This method may result in a small subdiscipline being merged into an adjacent large subdiscipline even if they are different.

Our approach easily expands to any data set which has keywords such as reports and patents. For instance, we can easily detect the trend of ongoing research work and technology from reports and patent data. Furthermore, leveraging the advance of natural language processing techniques, we can utilize more abundant information from the text. Using this rich information, we expect to expand our approach to a more diverse dataset with no explicit keywords.

\begin{acknowledgments}

This work was motivated by the research project on the classification of the topics in physics, which was done by the Korean Physical Society and supported by the National Research Foundation. This research was supported by the MSIT(Ministry of Science and ICT), Korea, under the ITRC(Information Technology Research Center) support program(IITP-2020-2018-0-01441) supervised by the IITP(Institute for Information \& Communications Technology Planning \& Evaluation), and partly supported by IITP grant funded by the Korea government(MSIT) (No.2019-0-01906, Artificial Intelligence Graduate School Program(POSTECH)) and Korea Evaluation Institute of Industrial Technology(KEIT) grant funded by the Korea government(MOTIE).

\end{acknowledgments}

\end{document}